\documentclass[submission,copyright,creativecommons]{eptcs}
\providecommand{\event}{ICE 2020}

\usepackage{latexsym,amssymb,url,boxedminipage}
\usepackage{amsmath}
\usepackage{amssymb}
 \usepackage{times}
 \usepackage{enumerate}
\usepackage{listings, color}
\usepackage{xspace}
\usepackage[final]{graphicx}
\usepackage{float}
\usepackage[caption = false]{subfig}

\title{Microservice Interface Based Deployment Orchestration\thanks{Research partly supported by the H2020-MSCA-RISE project ID 778233 ``Behavioural Application Program Interfaces (BEHAPI)''.}}
\author{Lorenzo Bacchiani
\institute{University of Bologna, Italy}
\email{lorenzo.bacchiani@studio.unibo.it}
\and
Mario Bravetti
\institute{University of Bologna, Italy / INRIA, France}
\email{mario.bravetti@unibo.it}
\and 
Saverio Giallorenzo
\institute{University of Bologna, Italy}
\email{saverio.giallorenzo@gmail.com}
\and 
Jacopo Mauro
\institute{University of Southern Denmark, Denmark}
\email{mauro.jacopo@gmail.com}
\and
Iacopo Talevi
\institute{University of Bologna, Italy}
\email{iacopo.talevi@gmail.com}
\and
Gianluigi Zavattaro
\institute{University of Bologna - Italy/ INRIA - France}
\email{gianluigi.zavattaro@unibo.it}
}

\def\titlerunning{Microservice Interface Based Deployment Orchestration}
\def\authorrunning{L. Bacchiani, M. Bravetti, S. Giallorenzo, J. Mauro, I. Talevi, G. Zavattaro}

\begin{document}

\long \def\ignore#1{\relax}

\newcommand{\mysql}{MySQL}

\newcommand{\Nodes}{\ensuremath{\mathcal{N}}}
\newcommand{\GetNode}{\ensuremath{.\mymathtt{node}}}
\newcommand{\GetConf}{\ensuremath{.\mymathtt{conf}}}
\newcommand{\GetCost}{\ensuremath{.\mymathtt{cost}}}
\newcommand\fun{\rightarrow}
\newcommand\dom{\mymathtt{dom}}
\newcommand\ran{\mymathtt{ran}}

\newcommand{\ParNames}{\ensuremath{\mathcal{N}}}
\newcommand{\Components}{\Resources}
\newcommand{\Locations}{\ensuremath{\mathcal{L}}}
\newcommand{\CompTy}[1]{\ensuremath{\mathcal{#1}}}
\newcommand{\Plan}[1]{\ensuremath{\mathsf{#1}}}
\newcommand{\Binding}[4]{\ensuremath{\{[#1,#2], [#3,#4]\}}}
\newcommand{\diamondarr}{$\diamondsuit$-arrow}
\newcommand{\target}[2]{\ensuremath{\langle \ResTy{#1}_t,#2_t \rangle}}
\newcommand{\state}[1]{{\sf #1}}

\newcommand{\configuration}{Configuration\xspace}
\newcommand{\generation}{Generation\xspace}
\newcommand{\planning}{Planning\xspace}

\newcommand{\PKG}[1]{\texttt{#1}}
\newcommand{\FOSS}{FOSS}
\newcommand{\MANCOOSI}{Mancoosi}
\newcommand{\AEOLUS}{Aeolus\xspace}
\newcommand{\AEOLUSCORE}{{\AEOLUS} core}
\newcommand{\AEOLUSMINUS}{\AEOLUS$^{-}$\xspace}

\newcommand{\TODO}[1]{{\color{red}#1}}


\newcommand{\trans}[1]{\ensuremath{\xrightarrow{#1}}}
\newcommand{\sat}[3]{\ensuremath{#1 \models (#2,#3)}}
\newcommand{\satr}[4]{\ensuremath{#1 \models_{req} (#2,#3,#4)}}
\newcommand{\satp}[4]{\ensuremath{#1 \models_{prov} (#2,#3,#4)}}
\newcommand{\rcorrect}[1]{\ensuremath{\mathfrak{C}(#1)}}
\newcommand{\rscorrect}[1]{\ensuremath{\mathfrak{SC}(#1)}}
\newcommand{\correct}[2]{\ensuremath{\mathfrak{C}(#1,#2)}}
\newcommand{\scorrect}[2]{\ensuremath{\mathfrak{SC}(#1,#2)}}

\newcommand{\Ifaces}{\ensuremath{\mathcal{I}}}
\newcommand{\Resources}{\ensuremath{\mathcal{Z}}}
\newcommand{\Res}{\ensuremath{\mathcal{R}}}
\newcommand{\Actions}{\ensuremath{\mathcal{A}}}
\newcommand{\ResTypesF}{\ensuremath{{\Gamma}}}
\newcommand{\ResTy}[1]{\ensuremath{\mathcal{#1}}}
\newcommand{\Conf}[1]{\ensuremath{\mathcal{#1}}}
\newcommand{\GetTypeC}{\ensuremath{{.\mymathtt{type}}}}

\newcommand{\GetRes}[1]{\ensuremath{[#1]}}

\newcommand{\mymathtt}[1]{\mbox{\footnotesize\tt #1}}
\newcommand{\GetStates}{\ensuremath{.\mymathtt{states}}}
\newcommand{\GetInit}{\ensuremath{.\mymathtt{init}}}
\newcommand{\GetTrans}{\ensuremath{.\mymathtt{trans}}}
\newcommand{\GetProv}{\ensuremath{.\mymathtt{prov}}}
\newcommand{\GetReq}{\ensuremath{.\mymathtt{req}}}
\newcommand{\GetSReq}{\ensuremath{.\mymathtt{req$_{\mymathtt{s}}$}}}
\newcommand{\GetWReq}{\ensuremath{.\mymathtt{req$_{\mymathtt{w}}$}}}
\newcommand{\GetResource}{\ensuremath{.\mymathtt{res}}}
\newcommand{\GetProvMap}[1]{\ensuremath{.\mathbf{P}}(#1)}
\newcommand{\GetWReqMap}[1]{\ensuremath{.\mathbf{R}_w\mymathtt{map}}(#1)}
\newcommand{\GetSReqMap}[1]{\ensuremath{.\mathbf{R}}(#1)}
\newcommand{\GetTy}{\ensuremath{.\mymathtt{type}}}
\newcommand{\GetState}{\ensuremath{.\mymathtt{state}}}

\newcommand{\TRUE}{\textbf{true}}
\newcommand{\FALSE}{\textbf{false}}

\newcommand\pfun{\mathrel{\ooalign{\hfil$\mapstochar\mkern5mu$\hfil\cr$\to$\cr}}}
\newcommand{\sem}[1]{[\!\![ #1 ]\!\!]}
\newcommand{\numins}[3]{#1^{\#}_{\langle#2,#3\rangle}}
\newcommand{\numport}[2]{#1^{\#}_{\langle#2\rangle}}
\newcommand{\multiset}[1]{\{\!\!\{ #1 \}\!\!\}}

\newcommand{\AbsConf}[1]{\ensuremath{\mathcal{#1}}}
\newcommand{\Concretization}[1]{\gamma(#1)}
\newcommand{\num}[2]{\#_{#1}(#2)}
\newcommand{\CorrectConf}{\mathit{Conf}}

\newcommand{\preset}[1]{^\bullet#1}
\newcommand{\postset}[1]{#1^\bullet}
\newcommand{\derivv}{\Rightarrow}
\newcommand{\resourceTrans}[1]{\eta( #1 )}
\newcommand{\resourceTransNoArg}{\eta\xspace}

\newcommand{\ResTyStatePair}[2]{\ensuremath{\langle \ResTy{#1},#2 \rangle}}
\newcommand{\pair}[2]{\ensuremath{\langle #1,#2 \rangle}}
\newcommand{\GetProvQ}[1]{\ensuremath{.\mathbf{P}}(#1)}
\newcommand{\GetReqQ}[1]{\ensuremath{.\mathbf{R}}(#1)}
\newcommand{\predarrow}{\ensuremath{\longrightarrow}}
\newcommand{\cparrow}{\hdashrule[0.5ex]{0.5cm}{1pt}{1pt}}
\newcommand{\add}{\texttt{add}}

\newcommand{\tuple}[1]{\langle #1 \rangle}
\newcommand{\nat}{\mathbb{N}}

\newcommand{\tooltxt}{SmartDepl\xspace}
\newcommand{\tool}{{\sf \tooltxt}\xspace}

\newcommand*\justify{%
  \fontdimen2\font=0.4em
  \fontdimen3\font=0.2em
  \fontdimen4\font=0.1em
  \fontdimen7\font=0.1em
  \hyphenchar\font=`\-
}

\lstdefinestyle{ANTLR}{
    basicstyle=\footnotesize\ttfamily\color{black},%
    breaklines=true,
    moredelim=[s][\color{magenta}\ttfamily]{'}{'},
    moredelim=*[s][\color{black}\ttfamily]{options}{\}},
    commentstyle={\color{gray}\itshape},
    morecomment=[l]{//},
    emph={%
        INT,
        VARIABLE,
        ID,
        RE
        },emphstyle={\color{blue}\ttfamily},
    alsoletter={:,|,;},%
    morekeywords={:,|,;},
    keywordstyle={\color{black}},
}

\lstdefinestyle{json}{
    basicstyle=\footnotesize\ttfamily,
    showstringspaces=false,
    breaklines=true,
    moredelim=[s][\color{magenta}\ttfamily]{"}{"},
    morekeywords={:,[,],\{\}},
    literate=
     *{0}{{{\color{blue}0}}}{1}
      {1}{{{\color{blue}1}}}{1}
      {2}{{{\color{blue}2}}}{1}
      {3}{{{\color{blue}3}}}{1}
      {4}{{{\color{blue}4}}}{1}
      {5}{{{\color{blue}5}}}{1}
      {6}{{{\color{blue}6}}}{1}
      {7}{{{\color{blue}7}}}{1}
      {8}{{{\color{blue}8}}}{1}
      {9}{{{\color{blue}9}}}{1},
}

\lstdefinelanguage{ABS}{keywords=
{null,this,thisDC,dyndelta,new,data,type,def,case,of,cog,class,interface,extends
,implements,if,else,await,get,total,load,transfer,movecogto,Fut,return,skip,
while,module,duration,now,deadline,import, export, uses, from, destiny, 
suspend,delta,adds,modifies,removes,original,productline,features,core,
corefeatures,optionalfeatures,after,when,product,hasAttribute,hasMethod,root,
extension,group,allof,oneof,require,exclude,original,ifin,ifout,opt}, 
sensitive=true, comment=[l]{//}, morecomment=[s]{/*}{*/}, morestring=[b]"}

\lstdefinestyle{absstyle}{
language=ABS,
columns=fullflexible,
mathescape=true,%
showstringspaces=false,%
keywordstyle=\bf\sffamily,
commentstyle=\sl\sffamily,%
basicstyle=\footnotesize\sffamily,
inputencoding=latin1, 
literate=*,
extendedchars
}

\lstdefinelanguage{MYSPEC}{keywords=
{sum,and,or,impl,forall,in,of,type,used,by,exists}, 
sensitive=true,morecomment=[s]{'}{'}}

\lstdefinestyle{declangstyle}{
language=MYSPEC,
columns=fullflexible,
mathescape=true,%
showstringspaces=false,%
keywordstyle=\bf\ttfamily,
commentstyle=\sl\ttfamily\color{magenta},%
basicstyle=\footnotesize\ttfamily,
inputencoding=latin1,
literate=*,
extendedchars
}


\newcommand{\doi}[1]{\url{http://dx.doi.org/#1}}

\maketitle
\begin{abstract}
\vspace{-.4cm}
Following previous work on the automated deployment orchestration of 
component based applications, where orchestrations are expressed in terms of 
behaviours satisfying component interface functional dependences, 
we develop a formal model specifically tailored for microservice 
architectures. The first result that we obtain is 
decidability of the problem of synthesizing optimal deployment orchestrations 
for microservice architectures, a problem that is, instead, 
undecidable for generic 
component-based applications.  
We, thus, show how optimal deployment 
orchestrations can be synthesized and how, by using such orchestrations we can devise a procedure for run-time adaptation based on performing global reconfigurations. Finally, we evaluate the applicability of our approach on a realistic 
microservice architecture taken from the literature. In particular, we use the
high-level object-oriented probabilistic and timed process algebra Abstract Behavioural Specification (ABS) 
to model such a case study and to simulate it. The results of simulation show the advantages of global reconfiguration w.r.t.\ local adaptation.

\vspace{-.2cm}
\end{abstract}

\section{Introduction}

Inspired by service-oriented computing, microservices structure software
applications as highly modular and scalable compositions of fine-grained and
loosely-coupled services~\cite{DGLMMMS17}. These features support modern
software engineering practices, like continuous
delivery/deployment~\cite{continuous_delivery} and application
autoscaling~\cite{autoscaling}. A relevant problem in these practices consists
of the automated deployment of the microservice application, i.e.\ the
distribution of the fine-grained components over the available computing
nodes, and its dynamic modification to cope, e.g.\  with positive or negative
peaks of user requests. Although these practices are already beneficial, they can be further improved by exploiting the interdependencies within an architecture (interface functional dependences), instead of focusing on the single microservice. Indeed, architecture-level deployment orchestration can:
\begin{itemize}
\vspace{-.1cm}
\item Optimize global scaling - e.g., avoiding the overhead of redundantly detecting inbound traffic and sequentially scale each microservice in a pipeline.
\vspace{-.1cm}
\item Avoid ''domino'' effects due to unstructured scaling - e.g., cascading slowdowns or outages.
\vspace{-.1cm}
\end{itemize}


In the presented paper, we report results from \cite{chapter} 
and additional work
on modeling and simulation, using the probabilistic and timed process algebra Abstract Behavioural Specification (ABS), for a case study: a real-world microservice architecture, inspired by the email processing pipeline 
from Iron.io~\cite{ironIO}. The expressiveness of ABS allows us to devise a quite realistic and complex model of the case study. Moreover, the simulation shows effectiveness of the deployment 
orchestrations generated with the theory and tools in \cite{chapter}. In particular, the advantage of performing runtime adaptation via global system reconfigurations w.r.t.\ local component scaling.

\subsection{Summary of Results from \cite{chapter}}

\begin{figure}[h]
  \centering
  \includegraphics[scale = 0.43]{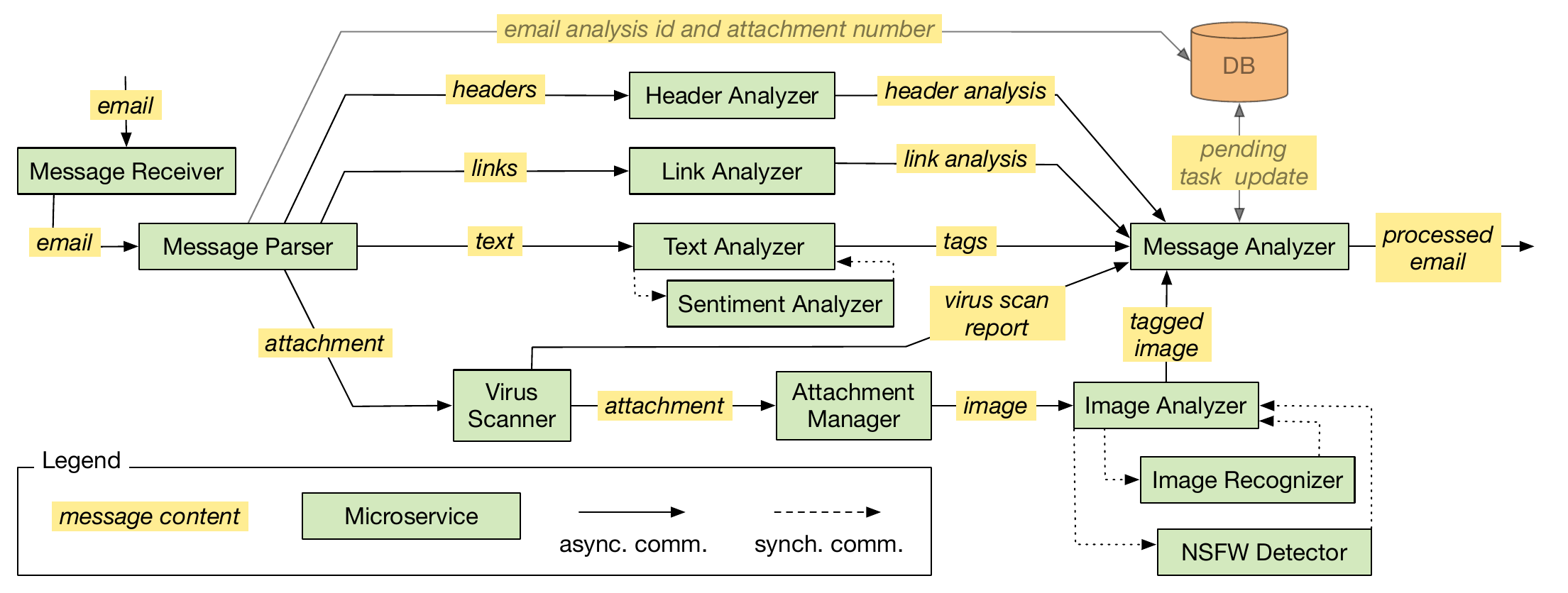}
  \caption{Microservice architecture for email processing pipeline.}
  \label{fig:arch}
\end{figure}

In \cite{chapter} we address the problem of orchestrating the deployment, and
re-deployment, of microservice architectures in a formal manner, by presenting
an approach for modeling microservice architectures, that allows us to both
prove formal properties and realize an implemented solution.
%
%
%
%
%
We follow the approach taken by the 
\emph{\AEOLUS component model}~\cite{sefm-aeolus,infCom14,concur15},
which was used to formally define the problem of
deploying component-based software systems and to prove that,
in the general case, such problem is undecidable~\cite{sefm-aeolus}.
The basic idea of \AEOLUS is to enrich the specification
of components with a finite state automaton 
that describes their deployment life cycle.
%

In \cite{chapter} 
we modify 
the \AEOLUS model 
in order to make it suitable for formal
reasoning on the deployment of microservices. To realize this purpose, we
significantly revisit the formalization of the deployment problem,
replacing old \AEOLUS components with a model of microservices.
The main difference between our microservices and \AEOLUS
components can be found in the composition of their deployment life cycle.
In lieu of using the total power of finite state automata,
as \AEOLUS and other TOSCA-compliant deployment models~\cite{Brogi15} do, 
we consider microservices to have two states: (i) creation and (ii)
binding/unbinding. About creation, we use \emph{strong} dependencies
to point out which microservices must be immediately connected to those just
created. After that, we use \emph{weak} dependencies to denote
which microservices can be
bound/unbound.
%
%
%
%
%
%
%
%
%
%
%
%
%
The rationale behind these changes comes from state-of-the-art
microservice deployment technologies like Docker~\cite{merkel2014docker} and
Kubernetes~\cite{Hightower}. In particular, we take the weak and strong dependencies
from
Docker Compose~\cite{docker_compose}, 
a language for defining multi-container Docker 
applications, that allows users to specify
different relationships among microservices 
using, e.g.\  the {\sf depends\_on} (resp. {\sf external\_links})
modalities that impose (resp.\@ do not impose) a specific startup
order, in the same way as our strong (resp.\@ weak) dependencies.
It is also convenient using weak dependencies to model horizontal scaling, e.g.\  a load
balancer 
bound to/unbound from many microservice instances during
its life cycle.
%
%

Moreover, w.r.t.\@ the \AEOLUS model,
we also take into account resource/cost-aware deployments, following
the {\sf memory} and {\sf CPU} resources found in Kubernetes.
The amount of resources microservices need to properly run, is 
directly added in their specifications. In a deployment, a system of microservices 
runs within a set of computation \emph{nodes}. 
Nodes represent
computational units, e.g.\  virtual machines in an Infrastructure-as-a-Service
Cloud deployment. Every node has a cost and a set of resources available to guest microservices.

On the model above, it is possible to introduce the \emph{optimal deployment problem} as follows: given
a starting microservice system, a set of available nodes, and a new target
microservice to be deployed, find a set of reconfiguration operations that,
once applied to the starting system, drives to a new deployment that includes the
target microservice. 
The abovementioned deployment is supposed to be \emph{optimal}, in the sense that the
overall cost, i.e.\ the sum of the costs, of the used nodes is is slightest.  
In \cite{chapter} we prove this problem to be
decidable~\cite{fase_paper} by
presenting an algorithm based on the generation of a set of constraints related to 
microservices distribution over nodes, connections to be established and
optimization metrics that minimize the total  
cost of the computed deployment. 
In particular, we investigate the possibility to actually solve
the deployment problem for microservices
by exploiting Zephyrus2~\cite{zephyrus2},
a configurator
optimizer that was originally envisaged for the Aeolus model~\cite{ase_paper_michael} but later extended and improved to support a new
specification language and the possibility to have preferences on the metrics
to optimize, e.g.\  minimize not only the cost but also the number
of microservices.

\subsection{Simulation with the Timed Process Algebra Abstract Behavioural Specification (ABS)}

\begin{figure}[h]
  \centering
  \includegraphics[scale=0.45]{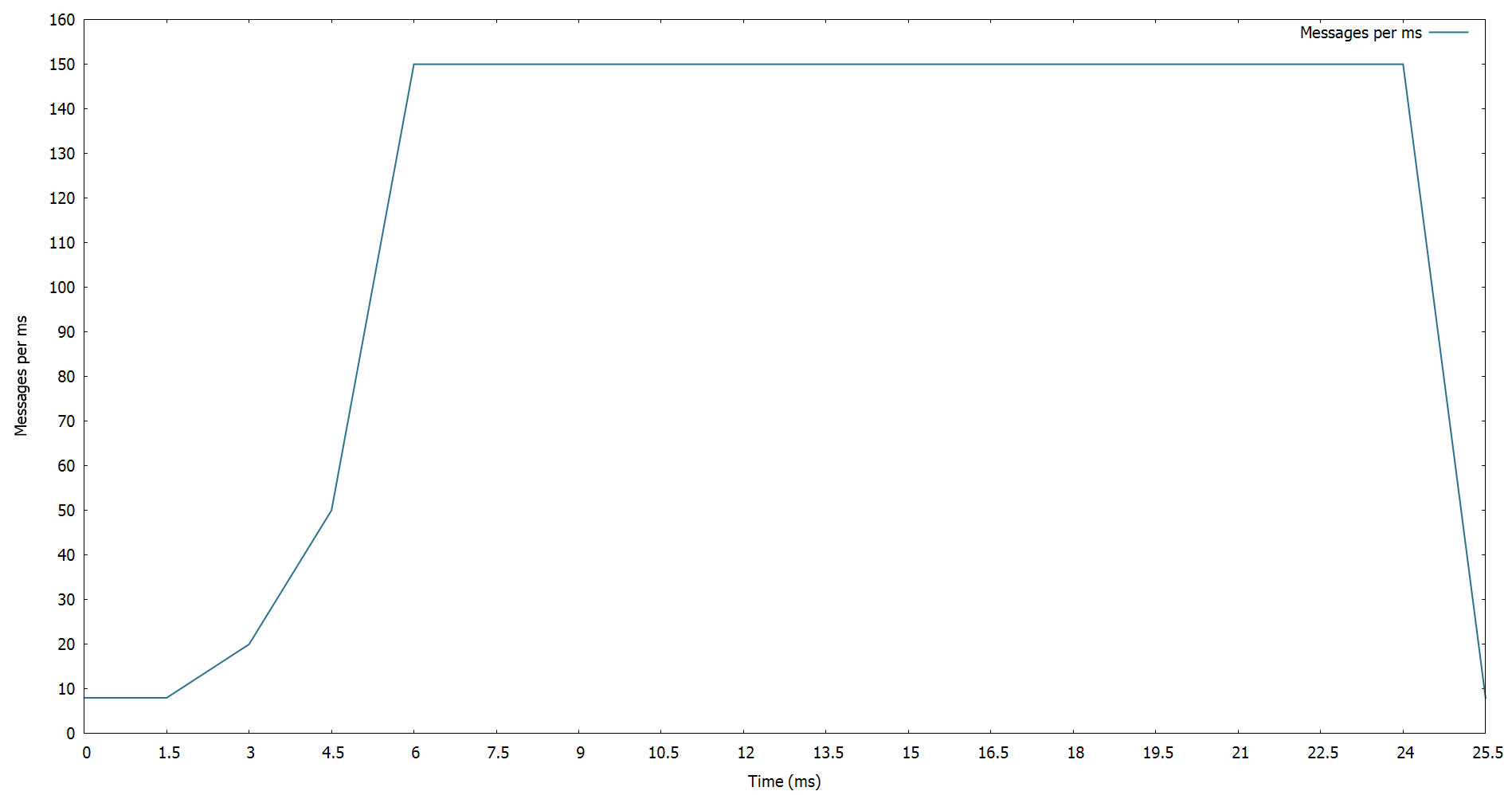}
  \caption{Simulated inbound frequencies.}
  \label{fig:freq}
\end{figure}

\begin{figure}[h]
  \centering
  \includegraphics[scale=0.45]{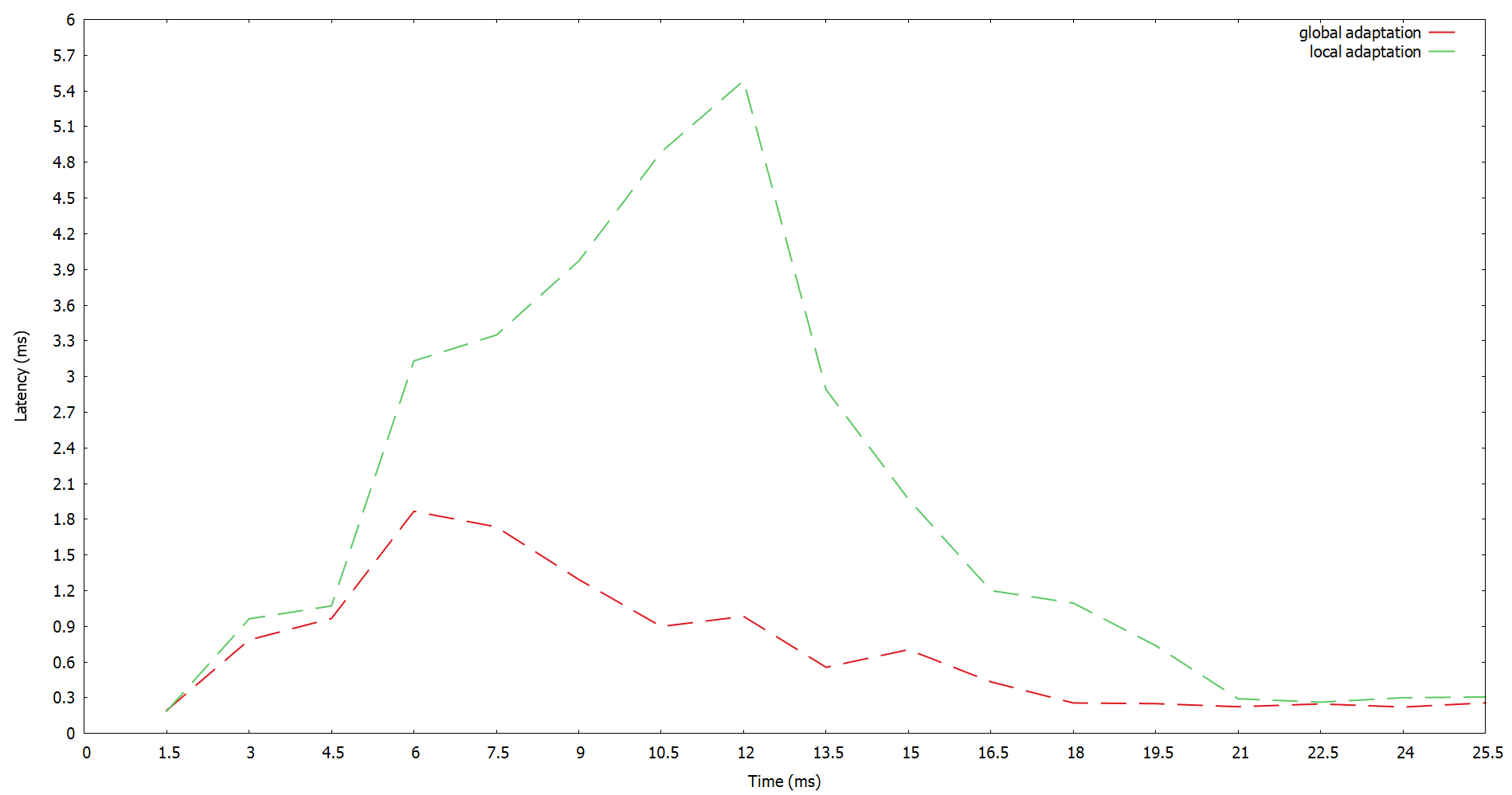}
  \caption{Latency comparison.}
  \label{fig:aat}
\end{figure}

\begin{figure}[h]
  \centering
  \includegraphics[scale=0.45]{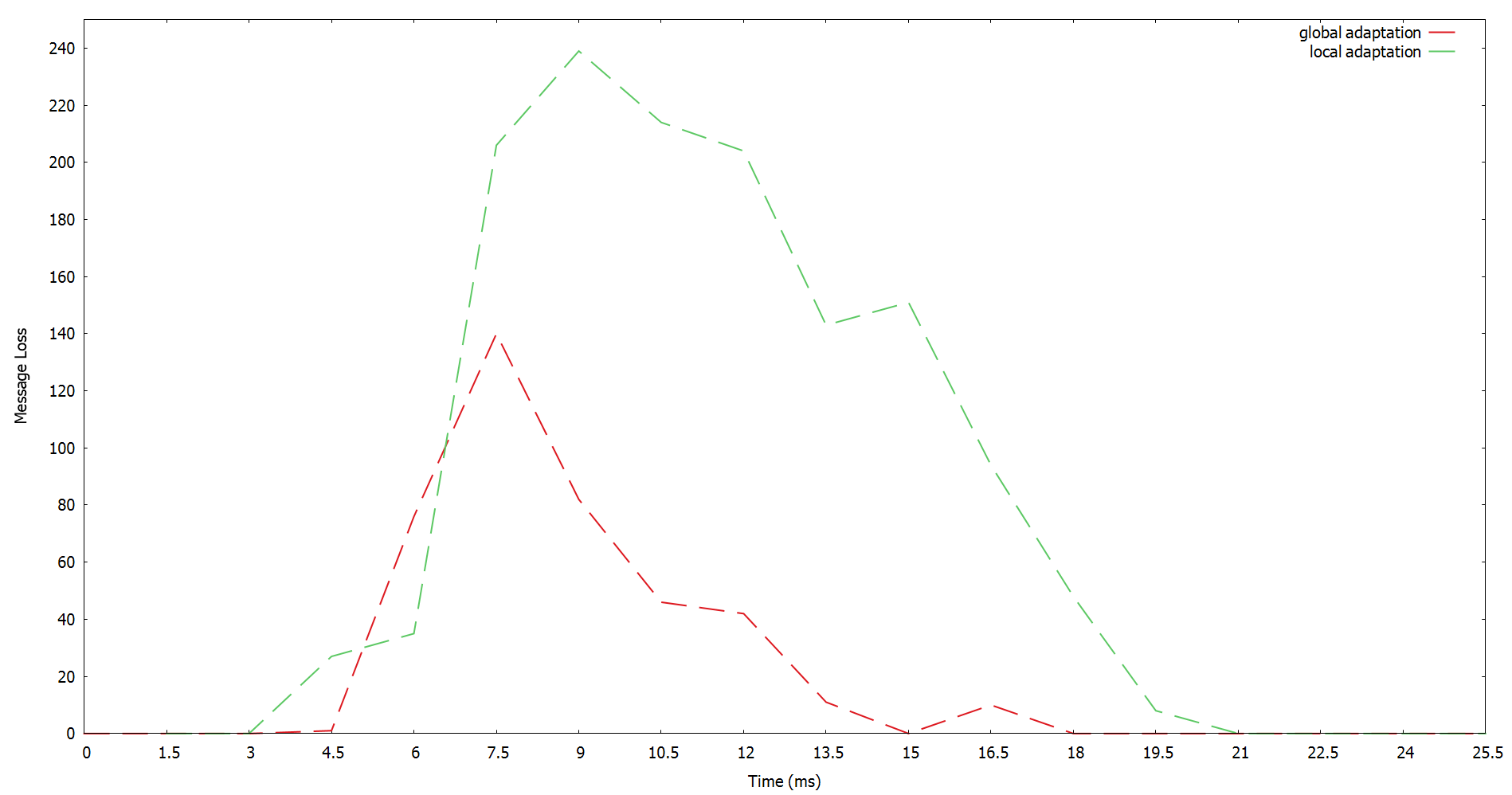}
  \caption{Loss comparison.}
  \label{fig:loss}
\end{figure}

\begin{figure}[h]
  \centering
  \includegraphics[scale=0.45]{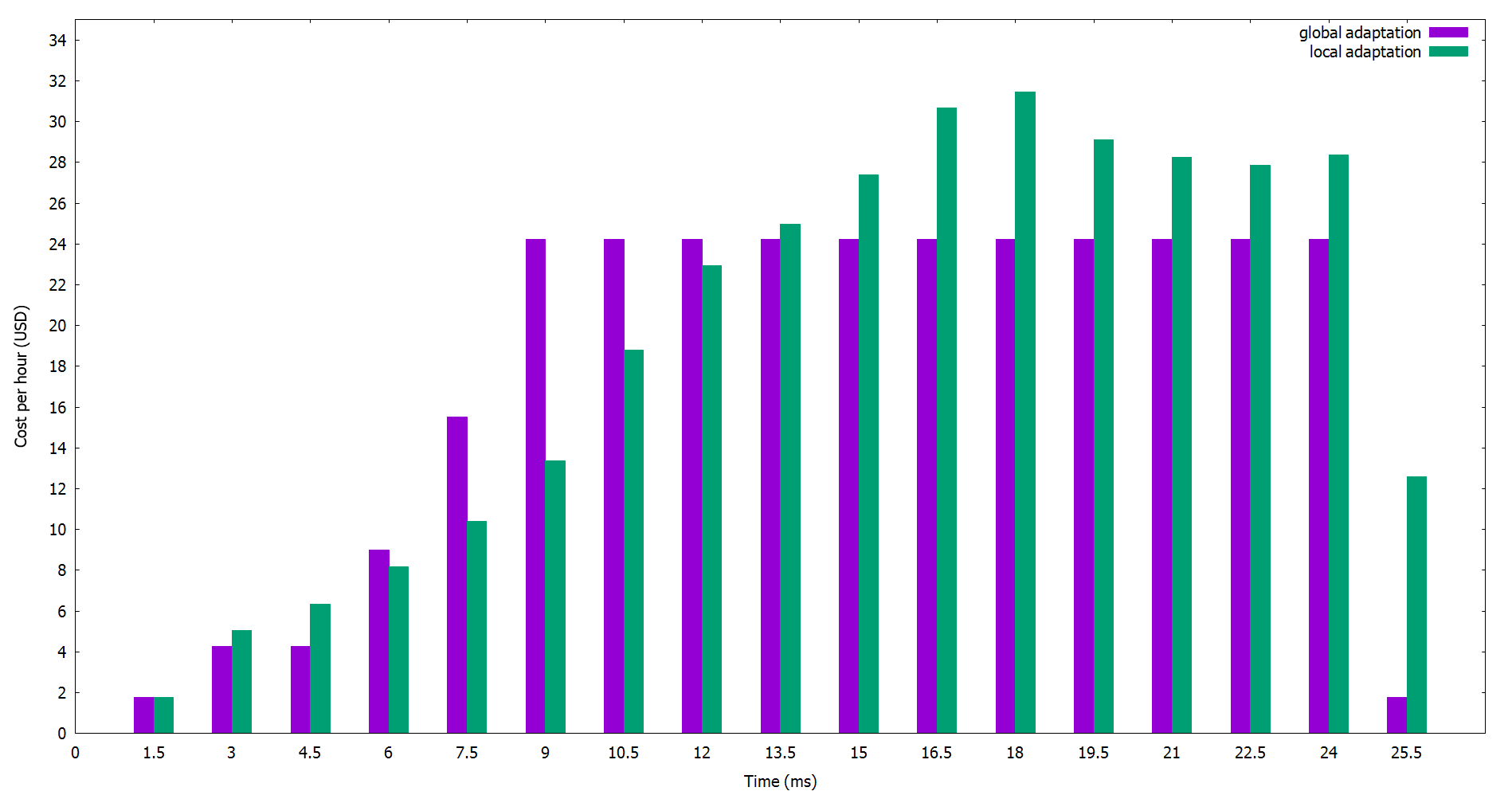}
  \caption{Costs comparison.}
  \label{fig:costs}
\end{figure}

\begin{figure}[h]
  \centering
  \includegraphics[scale=0.45]{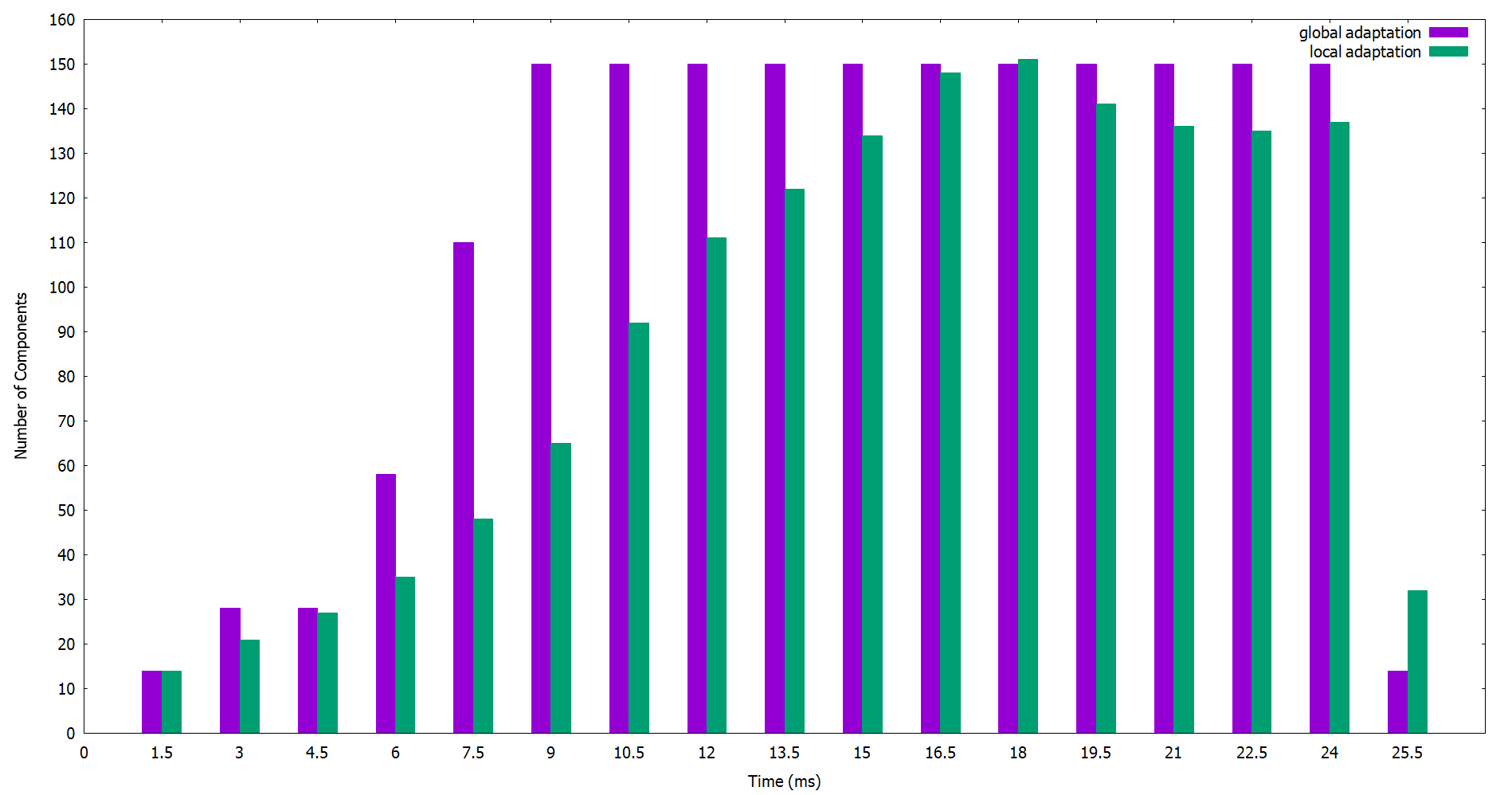}
  \caption{Number of components.}
  \label{fig:comp}
\end{figure}

We have evaluated the actual exploitability of the
approach of \cite{chapter}
by computing the initial optimal deployment, and run-time global
reconfigurations, for a real-world microservice architecture,
inspired by the reference email processing pipeline 
from Iron.io~\cite{ironIO}.
This architecture is modeled in the Abstract
Behavioral Specification (ABS) language, a high-level
object-oriented probabilistic and timed process algebra that supports asynchronous communication and deployment modeling~\cite{abs}.
Our technique is then used to compute two types of deployments: an initial one, with
one instance for each microservice, and a set of deployments to horizontally
scale the system depending on small, medium or large increments in the number
of emails to be processed. The experimental results 
are encouraging in that we were able to compute
deployment orchestrations for global reconfigurations that add, at each adaptation step, more than 30 new microservice instances, assuming
availability of hundreds of machines of three different types, and guaranteeing
optimality.

\subsubsection{The Email Pipeline Processing System}
The email processing pipeline described in~\cite{ironIO}, see Figure \ref{fig:arch} (taken from \cite{chapter}),
is composed by $12$ types of microservices and each type has its own load balancer. The architecture can be divided into four pipelines analyzing different parts of an email message. Messages enter the system through the \textit{Message Receiver} which forwards them to the \textit{Message Parser}. This component, in turn, extracts data from the email and routes them to a proper sub-pipeline. As expected the processing of each email component entails a specific working time. Each microservice can handle a specific workload, called max computational load - e.g., the \textit{Header Analyser} can handle a maximal inboud frequency of $40k$ requests per second, see \cite{chapter}. Finally, \cite{chapter} considers microservices to be deployed on Amazon EC2 virtual machines 
of type \textit{c4\_large}, \textit{c4\_xlarge} and \textit{c4\_2xlarge} respectively providing $2,4$ and $8$ cores. The case study architecture can be divided into four pipelines analyzing different parts of an email message. Messages enter the system through the \textit{Message Receiver}, which forwards them to the \textit{Message Parser}. This component, in turn, extracts data from the email and routes them to a proper sub-pipeline. Once each email component is processed, entailing a specific working time, analysis data is collected by the \textit{Message Analyzer} that produces an analysis report.

In the global adaptation approach scaling actions are provided by three reconfiguration orchestrations, i.e.\ \textit{Scale 1, Scale 2 and Scale 3}, which make system capable to deal with an augmented message inboud frequency ($+20k$, $+50k$ and $+80k$, respectively) w.r.t\ the maximum message workload in the base configuration: $10k$ messages per second, see \cite{chapter}. As we will show, these reconfiguration orchestrations minimize costs through the coexistence of microservices in the same computing node (virtual machine) and provide an architecture-level scaling making it possible to avoid cascading slowdowns.
The procedure governing the choice of the scaling orchestration is greedy. Indeed, taking the current message inbound frequency as a target, it computes the best scaling actions to apply based on minimizing the difference between the target value and the supported workload obtained by the scaling action under examination, until the system supports at least the target inbound frequency. 
After the target system configuration has been computed, the scaling actions required are executed and the system scales out.
On the contrary, the local adaptation approach simply scales every time a microservice constitutes a bottleneck by replicating it. As we will show in our simulation, this produces a chain effect, due also to time needed for deploying components at each step, which slows down the achievement of the target configuration necessary to handle the inbound messages frequency. Furthermore, each replica is hosted by a new node (instead of coexisting with other microservices) increasing costs more and more.

\subsubsection{System Modeling for Local and Global Adaptation}

Thanks to the expressiveness of the object-oriented probabilistic and timed process algebra ABS it was possible to model 
the email processing pipeline of Figure \ref{fig:arch}, including explicit modeling of load balancers, as ABS components/classes.
Each ABS component communicates asynchronously with other components (via {\it future} return types). Multiple ABS components are, themselves, located at a given deployment component, which is associated with a {\it speed} modeling its computational power: the number of computations per time unit it can perform. In our case study, an ABS time unit is set to model $.005$ milliseconds. 
As a matter of fact,  we built two ABS process algebraic models: one realizing the local adaptation mechanism discussed above and the other one implementing global reconfiguration via scaling actions \textit{Scale 1, Scale 2 and Scale 3}.
To make scaling operations effective, it is important to explicitly represent, within load balancers, request queues of a fixed maximal size. This explicit management not only provides a realistic model, but is also crucial for preventing the system from over-loading. Indeed, without these message queues, the system would refuse no messages and when the inbound workload grows up, it would overload the system with no possibility of restoring acceptable performances even if scaling actions occur. Moreover, queues allowed us to model message loss and to use it for comparing the behaviour of local and global scaling. 
Before our SmartDepl modification, the tool naively handled the {\it speed}: a total speed was statically assigned to a Deployment Component regardless of whether its cores were all used by actually deployed services or not. In timed ABS this caused microservices, deployed in a Deployment Component whose cores were not all used, to unrealistically proceed at a higher speed (as if they could exploit the computational power of the unused cores). Our solution is to statically set the Deployment Component {\it speed per core}, evaluate (in deployment orchestrations) the number of cores of a Deployment Component that are actually being used, and finally find the amount of speed ($speed\_per\_core \cdot unused\_cores$) to be subtracted to its total speed.
Moreover, we have additionally modeled real aspects such as the time that has to elapse before a service is ready to work.
We executed the two ABS process algebraic models (for local and global adaptation) by means of the Erlang backend provided 
as part of the ABS toolchain available at \cite{ABS_toolchain}. In order to build a complete simulation environment, we modeled (via an ABS data structure) a
message inbound frequency, see Figure \ref{fig:freq}, and the inner structure of messages via probabilistic contents  (exploiting the probabilistic features of ABS). As a matter of fact our two ABS models are, to the best of our knowledge, the biggest ones ever built with the ABS process algebra. Both ABS models are publicly available via GitHub at \cite{ABS_simulations}.

\subsubsection{Simulation Results}

As shown in Figure \ref{fig:freq} the simulated message inbound frequency grows rapidly until it reaches a stable situation so that we can test the adaptive responsiveness of the two approaches (local and global adaptation).

The first metric to be analyzed, in order to evaluate the performance of our new scaling approach, is the {\it latency}. We consider the latency as the average time for completely processing an email that enters the system. As it can be seen in Figure \ref{fig:aat}. Our approach, represented by the red dashed line, is extremely more performing than the classical one: considering that the maximum peak of incoming messages is reached around $6ms$, it restores a stable situation with a very low latency very fast. On the other hand, the classic approach, represented by the green dashed line, manages to reach an acceptable latency $16.5$ milliseconds after the peak. This is caused by the ability of the global adaptation strategy of detecting in advance the scaling needs of all system microservices. The above observation is confirmed by analyzing system {\it message loss}. Observing Figure \ref{fig:loss}, it is possible to see that our approach stops losing messages from $15$ milliseconds onwards, this means that message queues start to empty and latency can start to decrease.

{\it Costs} express how much a single virtual machine costs per hour. In Figure \ref{fig:costs}, we show a comparison between the sum of costs of all deployed virtual machines for both systems. The comparison shows, as we expected, that our approach is cheaper than the other. This is due to the fact that in the global adaptation approach components are placed into virtual machines using a constraint solver, \textit{Zephyrus2}, which minimizes an objective function (in this case, the cost).

Finally, comparing the {\it number of deployed components} helps to have a deeper understanding of the reasons why the global adaptation performs better. As it can be seen in Figure \ref{fig:comp} (where we also label the diagram with the structure of configurations in the case of global scaling), our approach reaches the target configuration needed to handle the maximum inbound workload faster than the classical approach and, obviously, this increments the adaptation responsiveness to higher workloads. The local adaptation slowness in reaching such a target configuration is caused by a scaling chain effect: local monitors check the workload every $10$ $sec$, so the number of components grows slowly 
due to the fact the microservices in the target configuration are not deployed together. For example, considering
the attachment pipeline in Figure \ref{fig:arch}, the first microservice to become a bottleneck is the \textit{Virus Scanner}: it starts losing messages, which will never arrive to the \textit{Attachment Manager}. Therefore, this component will not perceive the increment in the inbound emails until the \textit{Virus Scanner} will be replicated, thus causing a scaling chain effect that delays adaptation. This is the main cause for the large deterioration in performances observed in Figures \ref{fig:aat} and \ref{fig:loss}.

\newpage

\bibliographystyle{eptcs}
\bibliography{biblio}

\end{document}